\documentstyle[12pt,epsf]{article}

\begin{document}


\begin{center}

 {\Large \bf
\vskip 7cm
\mbox{EDDE Monte Carlo event generator.}
}
\vskip 1cm

\vskip 3cm

\mbox{Petrov~V.A., Ryutin~R.A., Sobol~A.E.}

\mbox{{\small Institute for High Energy Physics}}

\mbox{{\small{\it 142 281} Protvino, Russia}}

\mbox{and}

\mbox{Guillaud~J.-P.}

\mbox{{\small LAPP, Annecy, France}}

 \vskip 1.75cm
{\bf
\mbox{Abstract}}
  \vskip 0.3cm

\newlength{\qqq}
\settowidth{\qqq}{In the framework of the operator product  expansion, the quark mass dependence of}
\hfill
\noindent
\begin{minipage}{\qqq}

EDDE is a Monte Carlo event generator, under construction, for different Exclusive Double Diffractive 
Events. The program is based on the extended Regge-eikonal approach for "soft" 
processes. Standard Model and its extensions are used for "hard" fusion processes. An interface 
to PYTHIA, CMSJET and CMKIN is provided. 

\end{minipage}
\end{center}


\begin{center}
\vskip 0.5cm
{\bf
\mbox{Keywords}}
\vskip 0.3cm

\settowidth{\qqq}{In the framework of the operator product  expansion, the quark mass dependence of}
\hfill
\noindent
\begin{minipage}{\qqq}
Exclusive Double Diffractive Events -- Pomeron -- Regge-Eikonal model -- event generator
\end{minipage}

\end{center}

\setcounter{page}{1}
\newpage


\section{Introduction}

Usually High Energy Physics is considered as the synonym of "physics of particles". New phenomena in this area are related to discoveries of new particles or to the typical "particle-like" effects, such as "Bjorken scaling" in DIS or high-$p_T$ jets and so on. In the space-time language these kinematical regimes correspond to small distances.

However there is an area in High Energy Physics, which is related to rather large distances, even at high energies. These phenomena are similar to the scattering of hadrons to small angles. The wellknown feature of such processes is that the angle distribution gives the typical diffractive pattern with zero-angle maximum and one or, sometimes, two dips. Here wave properties of hadrons play the main role. From this distribution we can make the conclusion about size and shape of the scatterer, or the "interaction region".

Experiments at the LHC which aim to study low (TOTEM,…)- and high (CMS,…)-$p_T$ regimes, related to typical undulatory (diffractive) and corpuscular (point-like) behavior of the corresponding cross-sections, may offer a very exciting possibility to observe an interplay of both regimes. In theory the "hard part" can be (hopefully) treated with perturbative methods while the "soft" part is definitely nonperturbative.

Large number of event generators are devoted to partonic processes of the Standard Model and its extensions, i.e. work at small distances. It is wellknown, that perturbation theory has some problems in the description of processes at large distances. That is why diffractive processes are usually considered as special cases which description is based on different phenomenological approaches. The most popular approach is the Regge-eikonal model.

In the present paper a brife description of the generator EDDE (Exclusive Double Diffractive Events) is given. 


\section{Physics of EDDE.}

\begin{figure}[ht]
\label{pp_pXp}
\mbox{a)}
\vbox to 7cm {\hbox to 7cm{\epsfxsize=7cm\epsfysize=7cm\epsffile{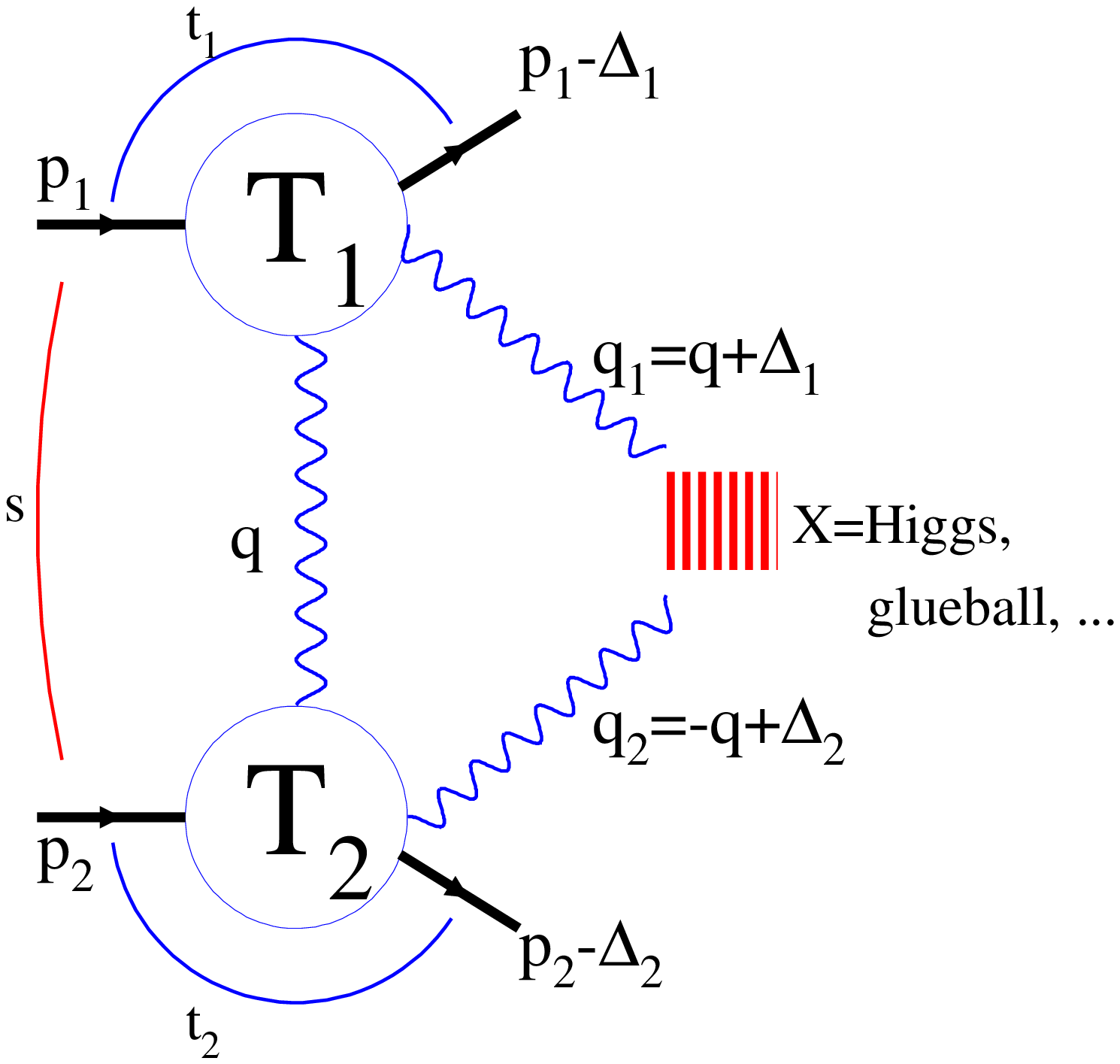}}}
\vskip -7cm
\hskip 8cm \mbox{b)}\vbox to 7cm {\hbox to 7cm{\epsfxsize=7cm\epsfysize=7cm\epsffile{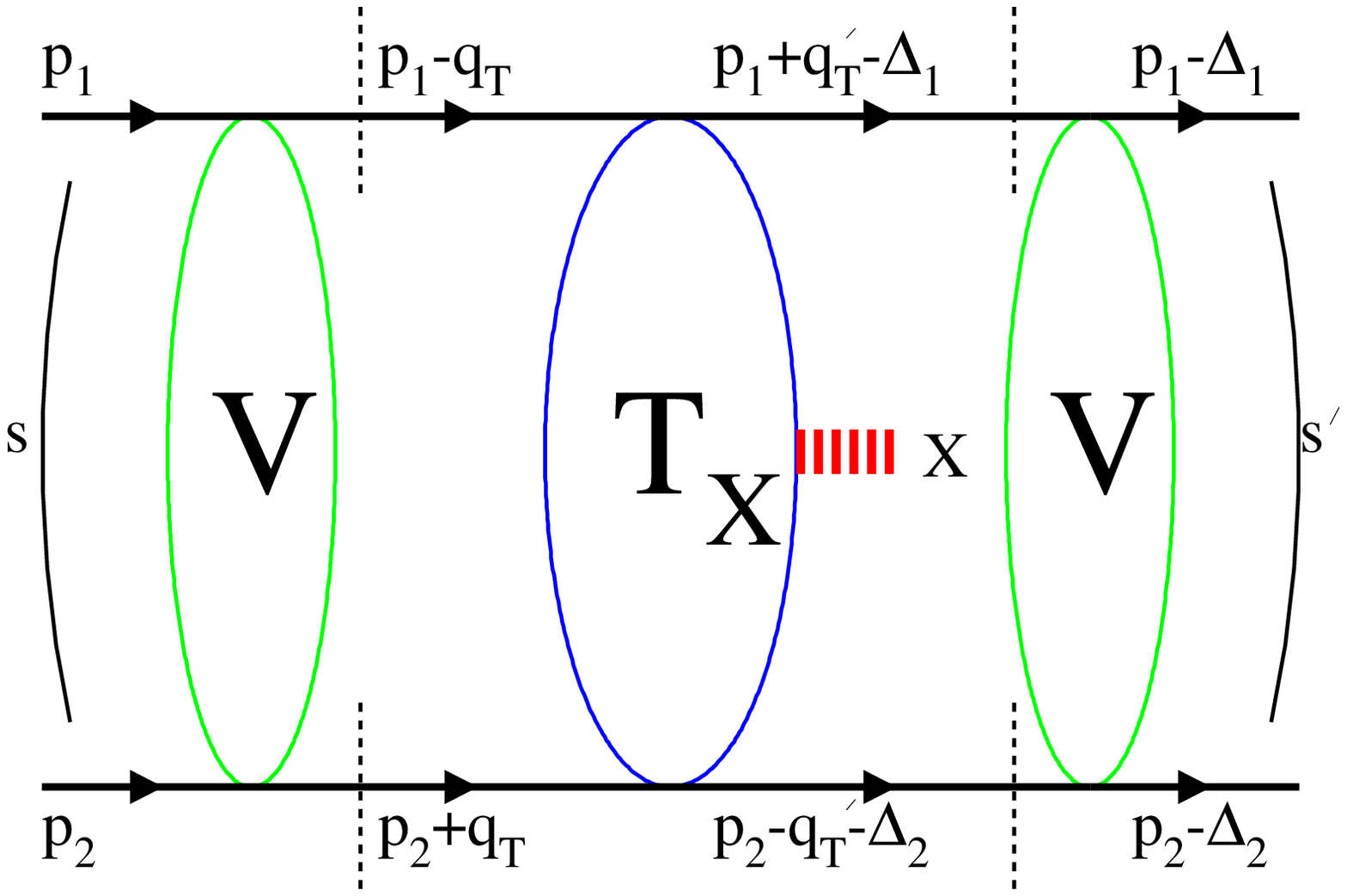}}}
\caption{a) The process $p+p\to p+X+p$. Absorbtion in the initial and final pp-channels is not shown. b) The full unitarization of the process $p+p\to p+X+p$.}
\end{figure}

For calculation of cross-sections we use the method developed in Refs.~\cite{1},\cite{2}. It is based on the extension of the Regge-eikonal approach, and succesfully used for the description of the HERA~\cite{3},\cite{4} and $p+p({\bar p})\to p+p({\bar p})$~\cite{5} data. 

Generator works in the kinematical region:
\begin{equation}
\label{tlimits}
0.01\; GeV^2\le |t_{1,2}|\le 1\; GeV^2\;{,} 
\end{equation}
\begin{equation}
\label{xlimits}
\xi_{min}\simeq\frac{M_X^2}{s}\le \xi_{1,2}\le \xi_{max}=0.1\;,
\end{equation} 

The amplitude of the process $p+p\to p+X+p$ can be obtained in the following way (see~Fig.~1). The first
step is to calculate the "bare" amplitude $T_X$, which is depicted in Fig.~1a. 

The "hard" part of $T_X$ is the usual gluon-gluon fusion process 
calculated by perturbative methods in the Standard Model or its extensions. In the first version EDDE1.1 three different processes are included:

\begin{itemize}
\item Standard model Higgs boson (H) production~\cite{H}. 
\item Randall-Sundrum model with one extra dimension~\cite{RS1}. Higgs boson (H$^*$) and Radion (R$^*$) production.
\item Standard model $b{\bar b}$ production.
\end{itemize}
For all the processes total cross sections and distributions are calculated.

The more recent version EDDE1.2 is available. There are some additional processes:
\begin{itemize}
\item Improved calculations for $b{\bar b}$ production, including  corrections to backround for Higgs or Radion production.
\item Standard model di-jet production.
\item Cross-sections for heavy quarkonia production at LHC.
\item Some azimuthal angular distributions.
\end{itemize}

"Soft" amplitudes $T_{1,2}$ are obtained 
in the extended Regge-eikonal approach. Parameters and formulaes can be found in Ref.~\cite{1}. The second step is the unitarization procedure, that takes into account initial and final state interactions (see Fig.~1b). 

\section{Program procedures and limits.}

There are several procedures and functions in the generator.

\underline{EDDE1.1}:
\begin{itemize}
\item function {\tt EDDECSHST(MH)} returns the total cross-section of Standard model Higgs boson production in fb. {\tt MH} is the Higgs boson mass in GeV. Typical values of the cross-section are $3.6\to 0.1$~fb in the mass region $100\to 500$~GeV.
\item function {\tt EDDECSRS1(XI,GAM,HMASS,RMASS,NP)} returns the total cross-section for H$^*$ ({\tt NP=1}) or R$^*$ ({\tt NP=2})production. {\tt XI} is the mixing parameter of the RS1 model~\cite{RS1}. Typical values are $-0.5\to 0.5$. {\tt GAM} is the scale parameter of the model, which is equal to $246\;{\rm GeV}/\Lambda$. $\Lambda$ is the Radion vacuum expectation value, which is directly related to the radius of compact extra-dimension. Usually $\Lambda=1\to 5\;{\rm TeV}$.
{\tt MASRS1(XI,GAM,HMASS,RMASS,NP)} returns the mass of the observable state H$^*$ ({\tt NP=1}) or R$^*$ ({\tt NP=2}). Input parameters are the same.
\item {\tt EDDECSBBBG(Mcut)} is the total cross-section for the leading order QCD $b{\bar b}$ production in the EDDE at LHC. {\tt Mcut} is the lower cut on the invariant mass of the $b{\bar b}$ system. {\tt Mcut}= 2 {\tt MTcut}, where {\tt MTcut} is the jet "transverse mass" cut, which is used in other subroutines. $\mbox{\tt MTcut}=\sqrt{E_{T, cut}^2+m_b^2}$ 
\item procedure {\tt EDDETTPHI(GT1,GT2,GPHI0)} generates the distribution
$$
\frac{1}{\sigma}\frac{d\sigma}{dt_1dt_2d\phi_0},
$$
where $\phi_0$ is the azimuthal angle between final protons.
\item $\xi$-distribution ($\xi=1-x_F$) is generated by the function {\tt EDDEX(HMASS)}, where {\tt HMASS} is the invariant mass of the central system.
\item procedure {\tt EDDEMXUBB(MTcut,UG,MXG)} generates the distribution for the "hard" process $gg^{PP}\to b{\bar b}$. {\tt MTcut} is the lower cut on the "transverse mass" of $b({\bar b})$ quark. 
\end{itemize}

\underline{EDDE1.2}:
\begin{itemize}
\item new procedure {\tt EDDECS0(NX,MTcut)} is introduced instead of {\tt EDDECSBBBG} which returns total cross-sections of $b{\bar b}$ or $gg$ di-jets and heavy quarkonia production at LHC.
$$
\mbox{\tt NX}=1\to b\;{\bar b}
$$
$$
\mbox{\tt NX}=2\to g\; g
$$
$$
\mbox{\tt NX}=3\to \chi_{c,0}
$$
$$
\mbox{\tt NX}=4\to \chi_{b,0}
$$
Parameter {\tt MTcut} is used for di-jet production only. In the STAGEN interface {\tt MTcut}={\tt Rpar(107)}. Default value is 25~ GeV.
\item procedure {\tt EDDEMXUGG(MTcut,UG,MXG)} generates the distribution for the "hard" process $gg^{PP}\to gg$. {\tt MTcut} is the lower cut on the transverse momentum of gluon (see above comment). 
\item additional modification of the procedure {\tt EDDETTPHI(GT1,GT2,GPHI0)} which is transformed to the more general one {\tt EDDETTPHI(NX,GT1,GT2,GPHI0)}, where {\tt NX} is the code number of produced particle or system of particles.
\begin{flushleft}
\begin{eqnarray}
&& \mbox{\tt NX}\nonumber\\
&& 1\to \mbox{Higgs, H*, R*, other heavy }0^{++}\;\mbox{states}\nonumber\\
&& 2\to b{\bar b}\nonumber\\
&& 3\to gg\nonumber\\
&& 4\to 0^{-+} {\rm state}\nonumber
\end{eqnarray}
\end{flushleft}
Other states will be added in the forthcoming version.
\end{itemize}

\section{Results.}

\begin{figure}[h]
\label{samples}
\mbox{a)}\hskip -0.5cm
\vbox to 7cm {\hbox to 7cm{\epsfxsize=7cm\epsfysize=7cm\epsffile{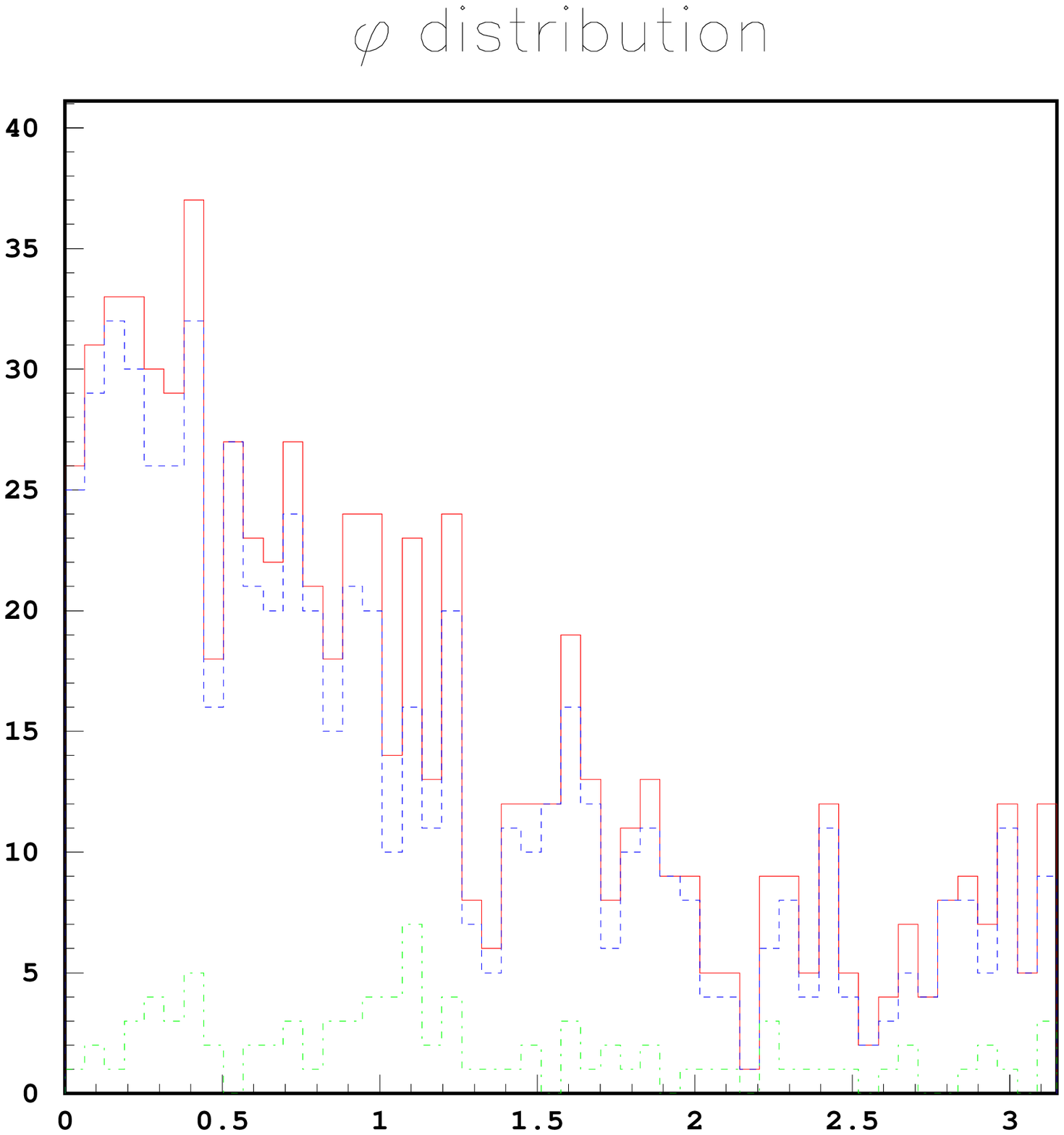}}}
\mbox{b)}
\vskip -7cm
\hskip 7cm
\vbox to 7cm {\hbox to 7cm{\epsfxsize=7cm\epsfysize=7cm\epsffile{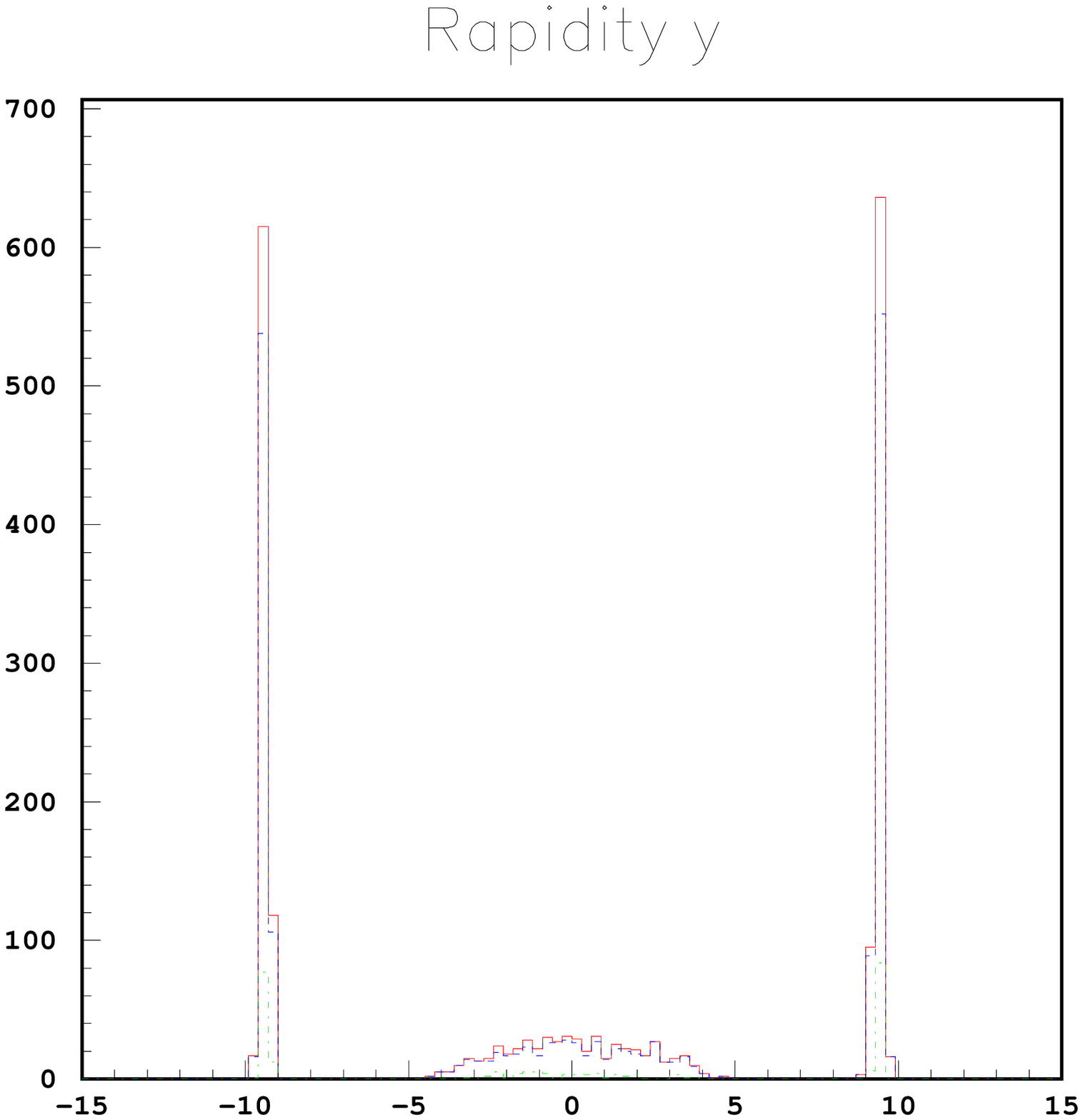}}}
\\
\mbox{c)}\hskip -0.5cm
\vbox to 7cm {\hbox to 7cm{\epsfxsize=7cm\epsfysize=7cm\epsffile{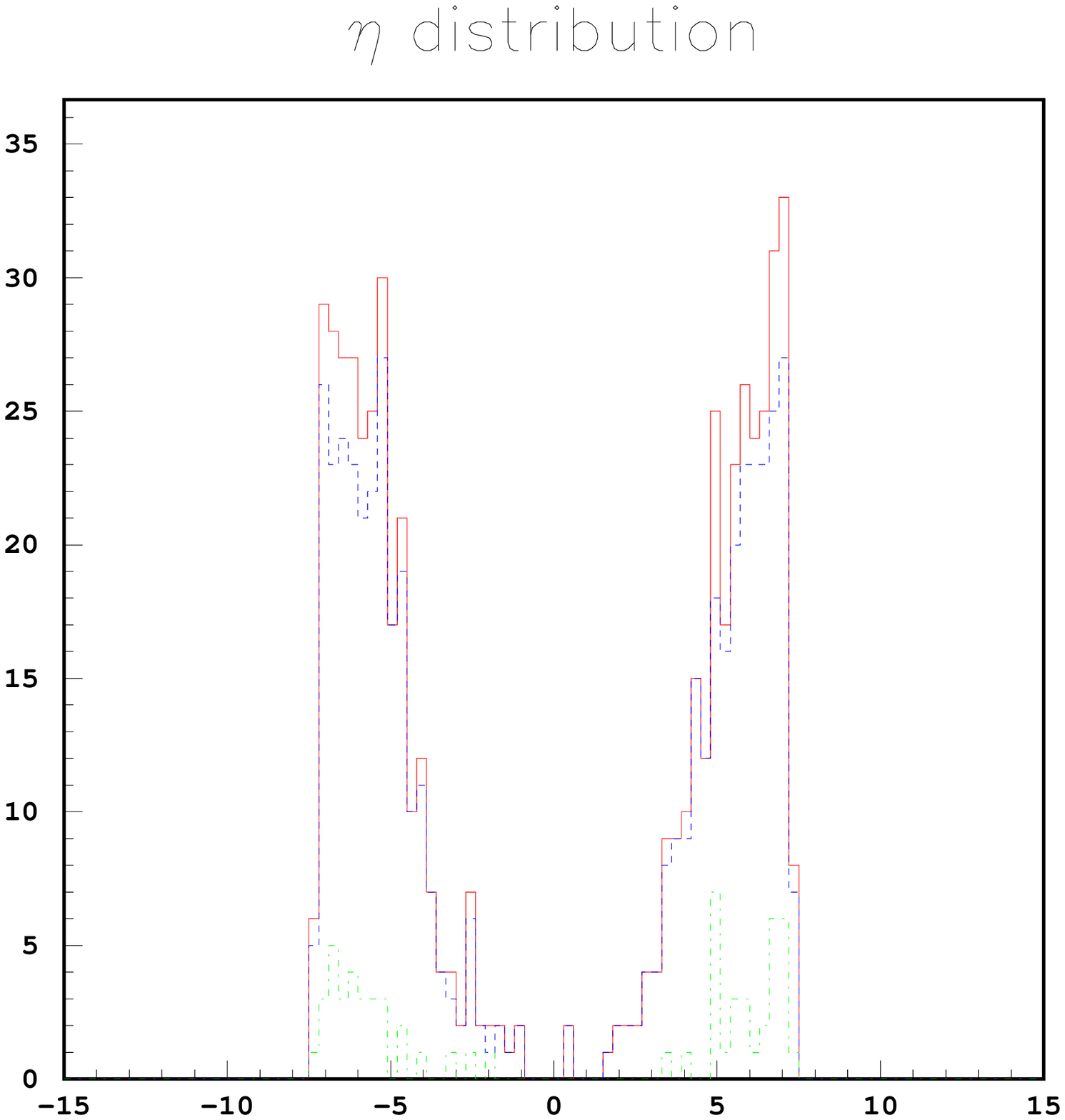}}}
\vskip -7cm
\hskip 7cm
\mbox{d)}
\vbox to 7cm {\hbox to 7cm{\epsfxsize=7cm\epsfysize=7cm\epsffile{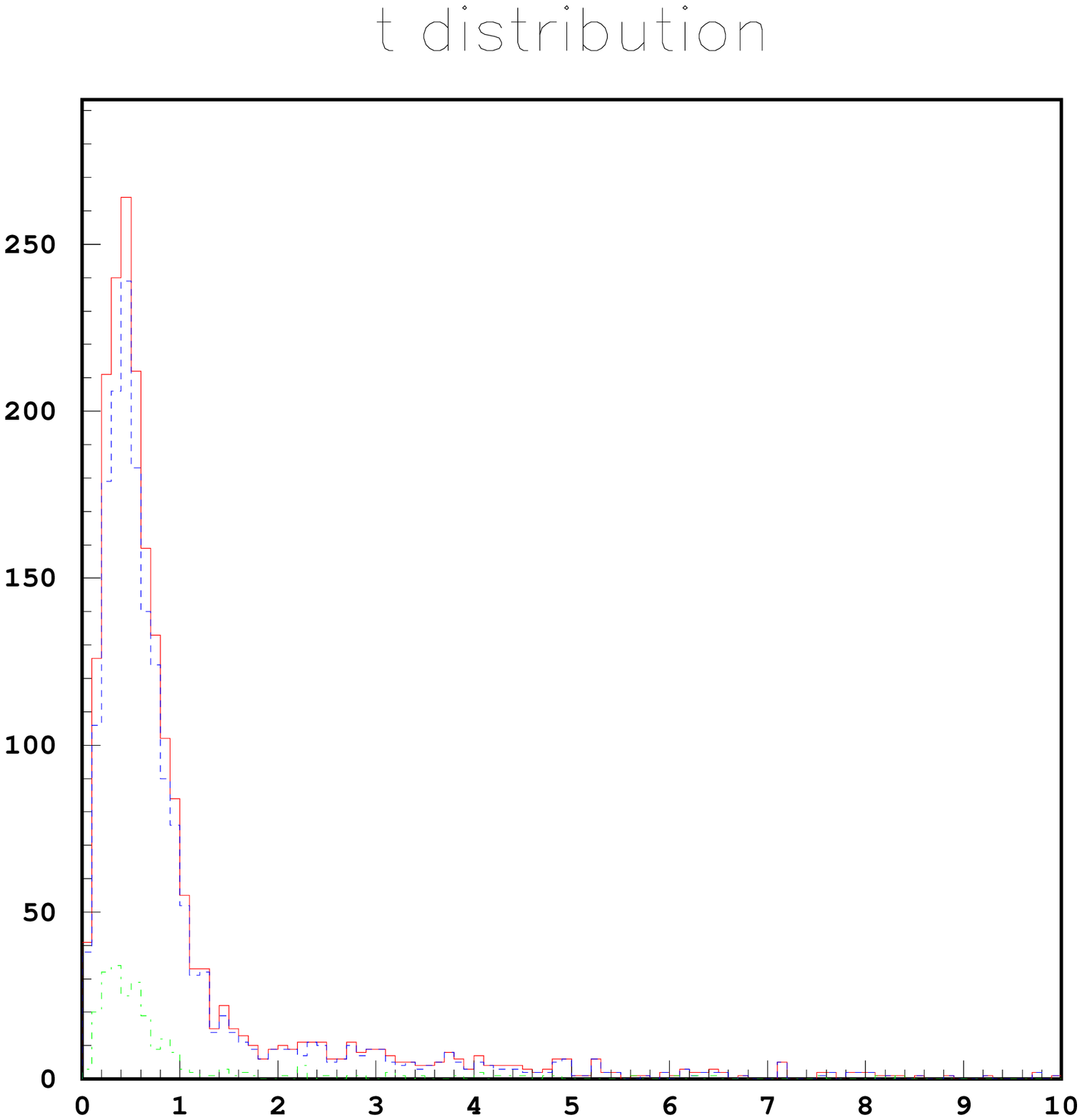}}}
\caption{Some distributions from EDDE1.2. (90 signal and 660 backround events. $MTcut=25$~GeV). Solid curve represent signal+background, dashed one is the background, and dash-dotted is the signal.  a) azimuthal angle distribution for Higgs boson production; b) rapidity and c) pseudorapidity distributions; d) integrated t-distributions.}
\end{figure}

Here some samples from the work of the generator with PYTHIA~\cite{pythia} are presented (see Figs.~3a)-d).) 

For both versions the iterface to the last version of PYTHIA is provided.

The updated version of the generator EDDE1.2 will be available on the web-page:
$$
http://sirius.ihep.su/cms/higgsdiff/diff.html
$$
Interface to CMKIN will be provided soon.

For the first version EDDE1.1 with interface to CMKIN and CMSJET~\cite{cmsjet} (in the STAGEN package) see the page
$$
http://cmsdoc.cern.ch/cms/generators/www/geners/collection/stagen/stagen.html
$$
Someone can use the new source file for STAGEN (see the webpage) to update EDDE1.1 until the new interface is provided.

\section*{Aknowledgements}

This work is supported by grants CNRS-PICS-2910 and RFBR-04-02-17299.

Thanks to Andrey Sobol for the first version of the generator {\tt DPEHiggs}, to 
Sergei Slabospitsky, who helped to correct the interface to PYTHIA and CMSJET, to Kreiton Hogg 
and Marek Tasevsky for helpful discussions concerning the work of the generator EDDE1.1.
 

\end{document}